\documentclass[11pt,a4paper]{article}
\usepackage[dvips]{graphicx}
\usepackage{amsmath,amssymb,latexsym}
\begin{document}
%
%
\pagestyle{headings}  
%
%
%
%
%
\title{Complexity Analysis in Cyclic Tag System Emulated by Rule 110}
%
%
\author{Shigeru Ninagawa \\
Kanazawa Institute of Technology, Ishikawa, Japan.\\
University of the West of England, Bristol, United Kingdom. \\
shigeruninagawa@gmail.com \\
\and Genaro J. Mart\'{i}nez \\
Departamento de Ciencias e Ingenier\'{i}a de la Computaci\'{o}n, \\
Escuela Superior de C\'{o}mputo, Instituto Polit\'{e}cnico Nacional,
M\'{e}xico, D. F.\\
University of the West of England, Bristol, United Kingdom. \\
genaro.martinez@uwe.ac.uk
}
%
%
%

\maketitle              

\begin{abstract}
It is known that elementary cellular automaton rule 110 is capable of supporting universal computation by emulating cyclic tag system. Since the whole information necessary to perform computation is stored in the configuration, it is reasonable to investigate  the complexity of configuration for the analysis of computing process. In this research we employed Lempel-Ziv complexity as a measure of complexity and calculated it during the evolution of emulating cyclic tag system by rule 110. As a result, we observed the stepwise decline of complexity during the evolution. That is caused by the transformation from table data to moving data and the elimination of table data by a rejector.
\end{abstract}
\section{Introduction}
Cellular automaton (CA) is a model of information processing system. Since CAs have no memory except for cell, all the information necessary to perform computation is stored in its configuration. That means the complexity of configuration is in some way related with the complexity of information the CA is processing. Therefore it is reasonable to investigate the complexity of configuration for the analysis of computing process by CA. Elementary CA (ECA) rule 110 is supporting universal computation~\cite{cook_04}. In this research we focus on the complexity of configurations during the computing process by rule 110. In the next section, we make a brief explanation of cyclic tag system emulated by rule 110. The results of complexity analysis of cyclic tag system are shown in section 3. Finally we discuss the results and a future plan.

\section{Cyclic Tag System by Rule 110}
The transition function of ECA rule 110 is given by:
\[
\frac{111}{0} \frac{110}{1} \frac{101}{1} \frac{100}{0} \frac{011}{1} \frac{010}{1}
\frac{001}{1} \frac{000}{0}.
\]
The upper line represents the state of the neighborhood and the lower line specifies the state of the cell at the next time step. Cook proved the computational universality of rule 110 by showing that rule 110 can emulate cyclic tag systems~\cite{cook_04}.

A cyclic tag system works on a finite tape which is read from the front and appended to based on what is read. An appendant is cyclically chosen from the appendant table. The alphabet on the tape consists of $\{ 0, 1 \}$. At each step, the system reads one character and deletes it, and if that is '1', then it appends the appendant, while an '0' causes the appendant to be skipped. At the next step, the system moves on to the next appendant in the table. The system halts if the word on the tape is empty.
For example, the transition of the initial word '1' with the appendant table (1, 101 ) is given as following:
\[
1 \vdash 1 \vdash 101 \vdash 011 \vdash 11 \vdash 11 \vdash 1101 \vdash \cdots.
\]
A detailed explanation of the emulation of cyclic tag systems by rule 110 is in Ref.~\cite{martinez_11}.

 \section{Complexity Analysis}
As a measure of complexity, we focus on the compressibility of configuration. Compression-based CA classification was implemented by Zenil~\cite{zenil_10} using the DEFLATE algorithm~\cite{deutsch_96}. 
We use Lempel-Ziv (LZ) complexity used in the data compression algorithm called LZ78~\cite{ziv_78}. In LZ78, a string is divided into phrases.  Given a string $s_1s_2 \cdots s_ks_{k+1} \cdots$ where a substring $s_1s_2 \cdots s_k$ has already been divided is constructed by searching the longest substring $s_{k+1} \cdots
s_{k+n} = w_j$, $(0 \le j \le m)$ and by setting $w_{m+1}=w_js_{k+n+1}$
where $w_0 = \epsilon$. The LZ complexity of the string is defined as the number of divided phrases. The complexity analysis based on LZ78 was applied to the study of the parity problem solving process by rule 60~\cite{ninagawa_13}.

\begin{figure}[t]]
\scalebox{0.73}{\includegraphics{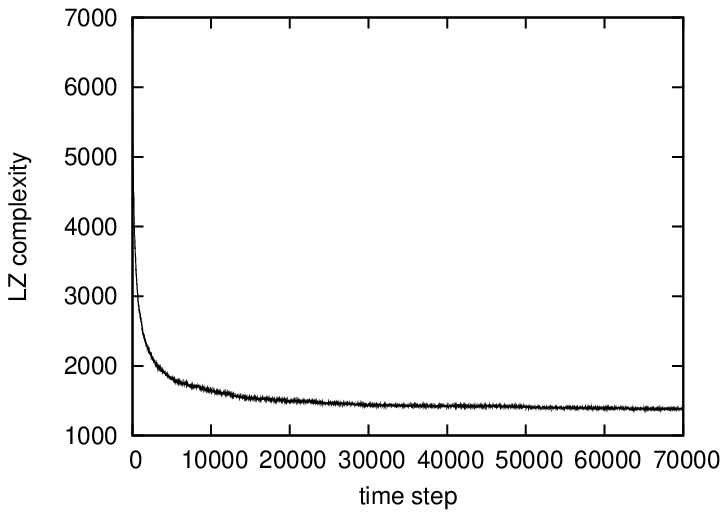}}
\scalebox{0.73}{\includegraphics{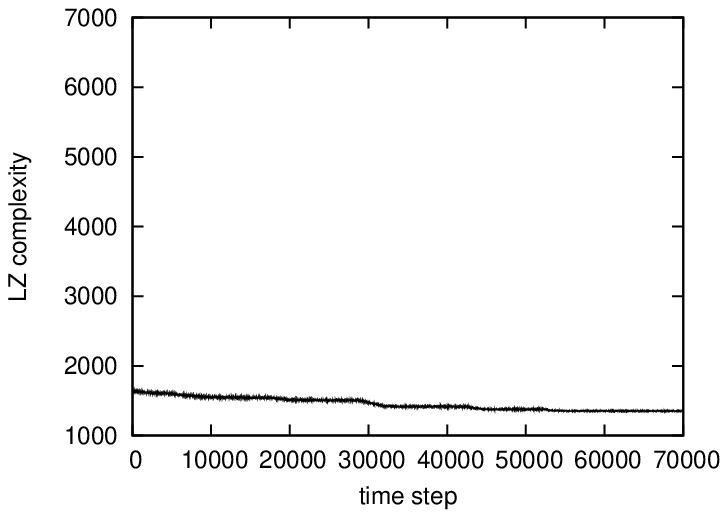}}
\caption{Evolution of LZ complexity staring from a random configuration (left) and from the one designed to emulate cyclic tag system (right).}
\label{LZ_evolution}
\end{figure}

In this research, we calculate the LZ complexity of configuration at each time step
during the computing process of cyclic tag system emulated by rule 110.
Figure~\ref{LZ_evolution} shows the evolution of LZ complexity staring
from a random configuration (left) and from the one designed to emulate
cyclic tag system. The array size is 65900 in both cases.
We made use of the initial configuration on the web site \cite{martinez_web} from which you can download the file of initial configuration of rule 110 emulating the cyclic tag system exemplified in the previous section.

\begin{figure}
\begin{center}
 \scalebox{0.8}{\includegraphics{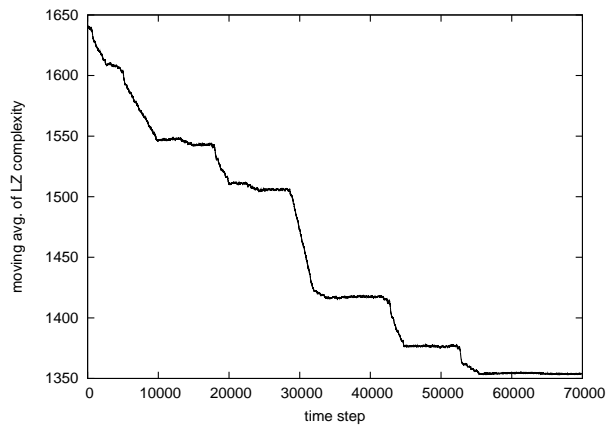}}
\caption{Moving average of LZ complexity in the evolution of cyclic tag system.}
\end{center}
 \label{moving_avg}
\end{figure}

\begin{figure}
\scalebox{0.8}{\includegraphics{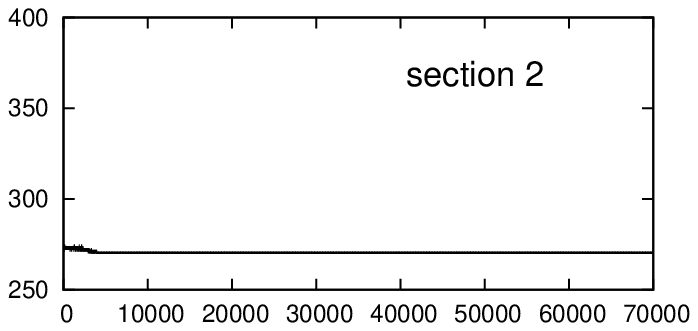}}
\scalebox{0.8}{\includegraphics{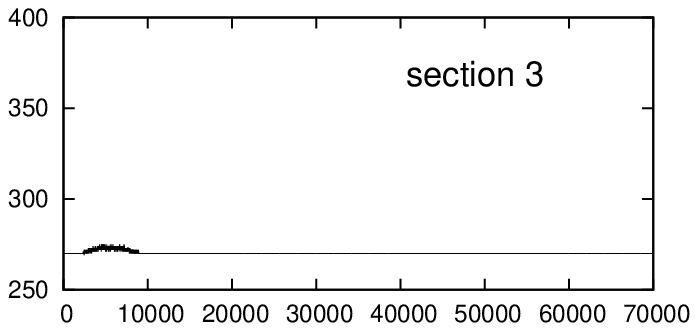}}
\scalebox{0.8}{\includegraphics{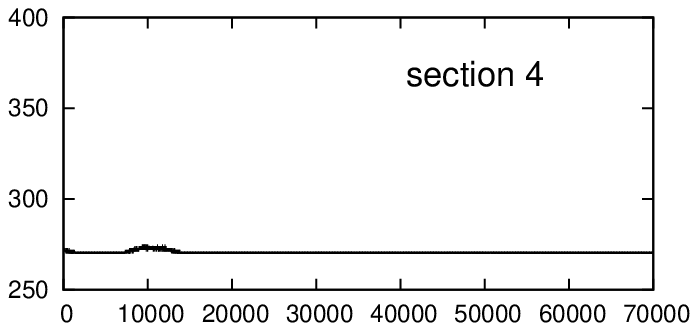}}
\scalebox{0.8}{\includegraphics{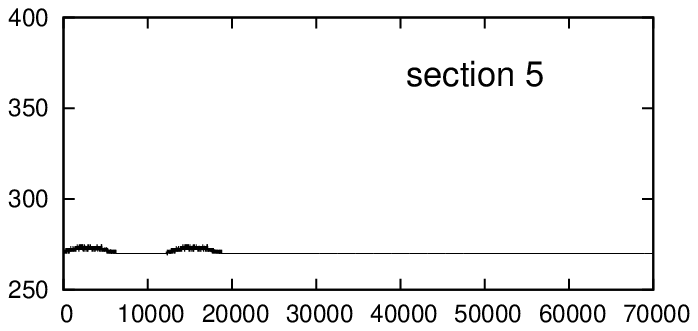}}
\scalebox{0.8}{\includegraphics{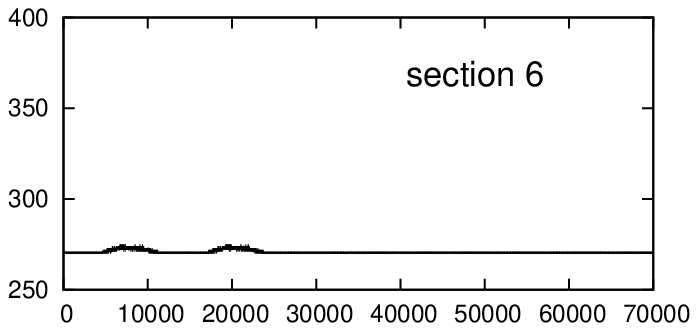}}
\scalebox{0.8}{\includegraphics{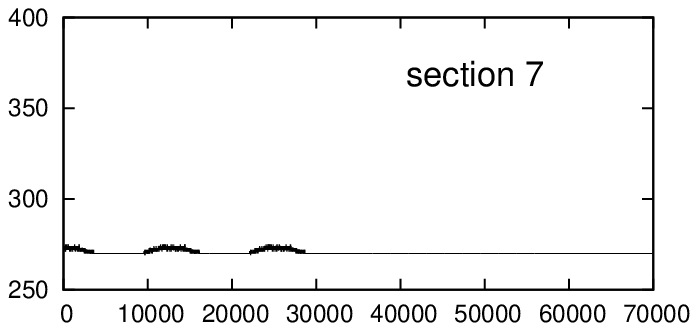}}
\scalebox{0.8}{\includegraphics{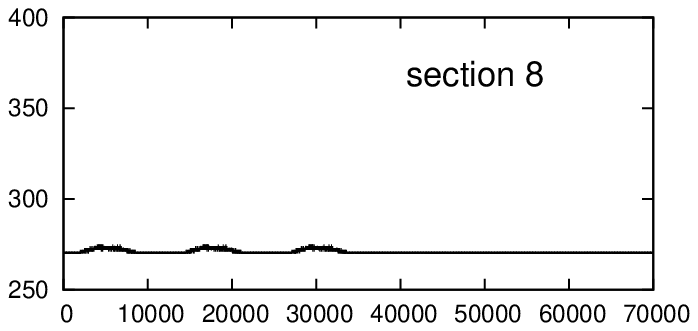}}
\scalebox{0.8}{\includegraphics{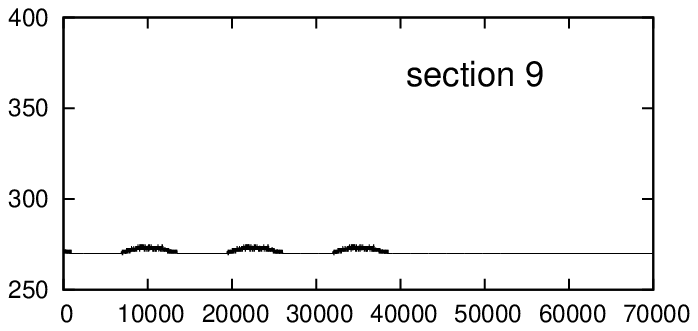}}
\scalebox{0.8}{\includegraphics{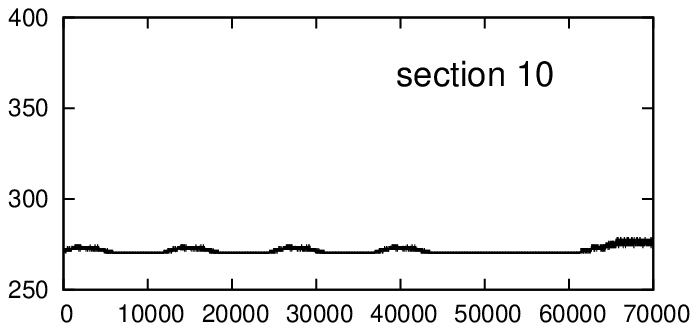}}
\scalebox{0.8}{\includegraphics{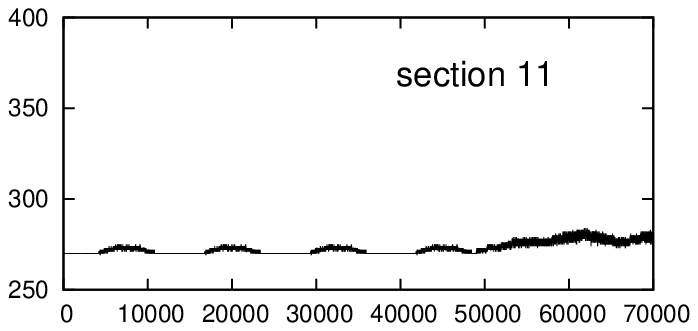}}
\scalebox{0.8}{\includegraphics{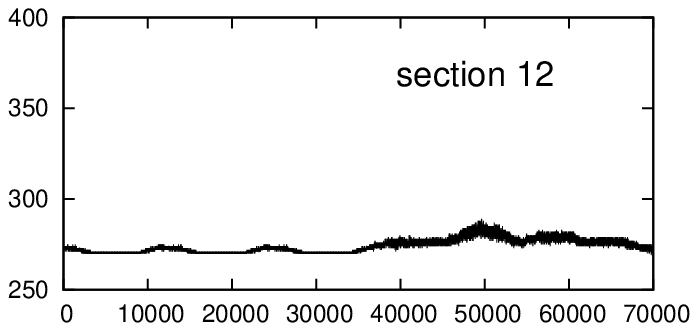}}
\scalebox{0.8}{\includegraphics{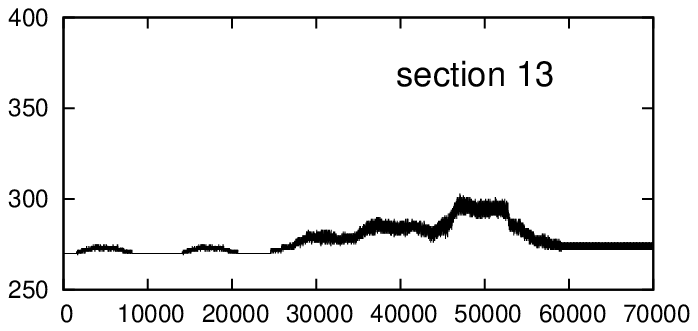}}
\scalebox{0.8}{\includegraphics{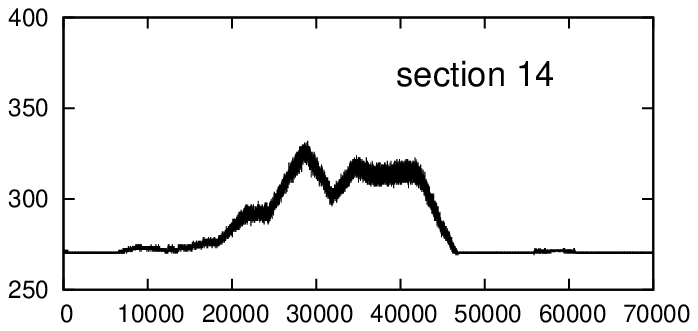}}
\scalebox{0.8}{\includegraphics{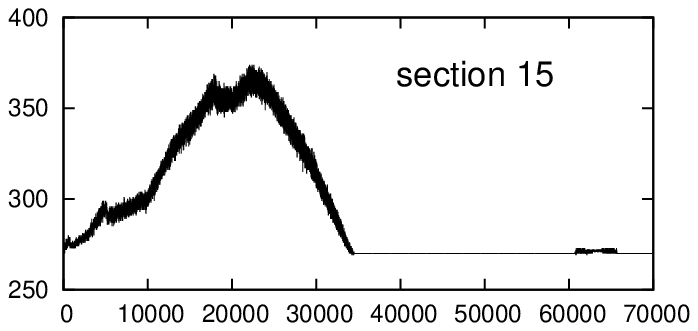}}
\scalebox{0.8}{\includegraphics{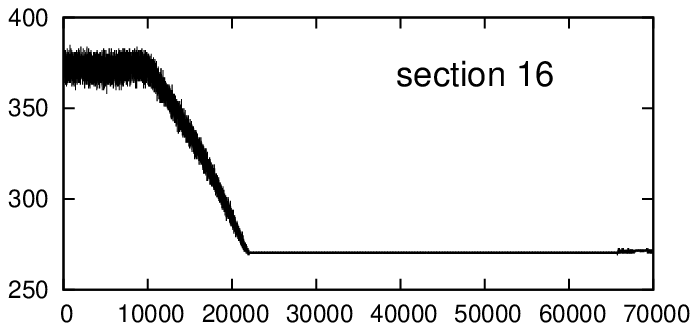}}
\scalebox{0.8}{\includegraphics{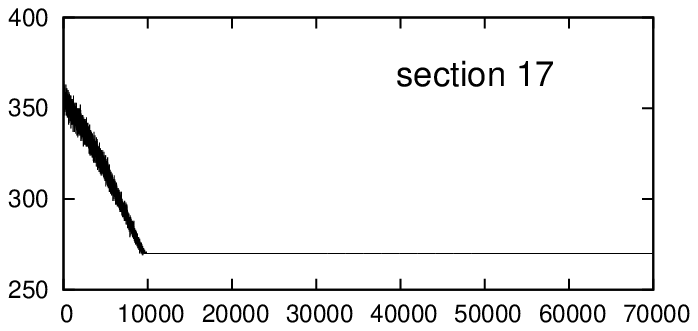}} 
\caption{Evolution of LZ complexity in section 2 - 17. Vertical axis is LZ complexity and horizontal axis is time step.}
\label{LZ_section_cont}
\end{figure}

In the case of a random initial configuration, the LZ complexity starts
with the value of 6068 and decreases
quickly, meanwhile, it shows a more slighter decrease in the case of cyclic tag system emulation.
To investigate the evolution of LZ complexity in the process of cyclic tag system emulation, we calculated the simple moving average of the data with period 100 and showed with a finer scale in Fig.~\ref{moving_avg}. We can see  the repetition of significant decline and temporary equilibrium in the evolution of LZ complexity. Figure~\ref{moving_avg}, however, does not inform us about  the regional difference of LZ complexity, because LZ complexity is a measure of complexity from a global perspective. So we divide the array into 20 sections  (section $0 \sim 19$ starting from the left) of 3295 cells and calculate the LZ complexity of each section individually. Figure~\ref{LZ_section_cont} shows the evolution of the LZ complexity in section $2 \sim 17$.

By viewing the figures from section 2 to 11 in sequence, we can observe four bumps are moving from the left to the right.
Each bump corresponds to the four packages of $A^4$ gliders that construct an appendant from a moving data.
Those packages of $A^4$ gliders are called ossifier ($4\_A^4$). The term in the parenthesis
is the one according to the naming convention employed in Ref~\cite{martinez_11}.

\begin{figure}[t]
\scalebox{0.34}{\includegraphics{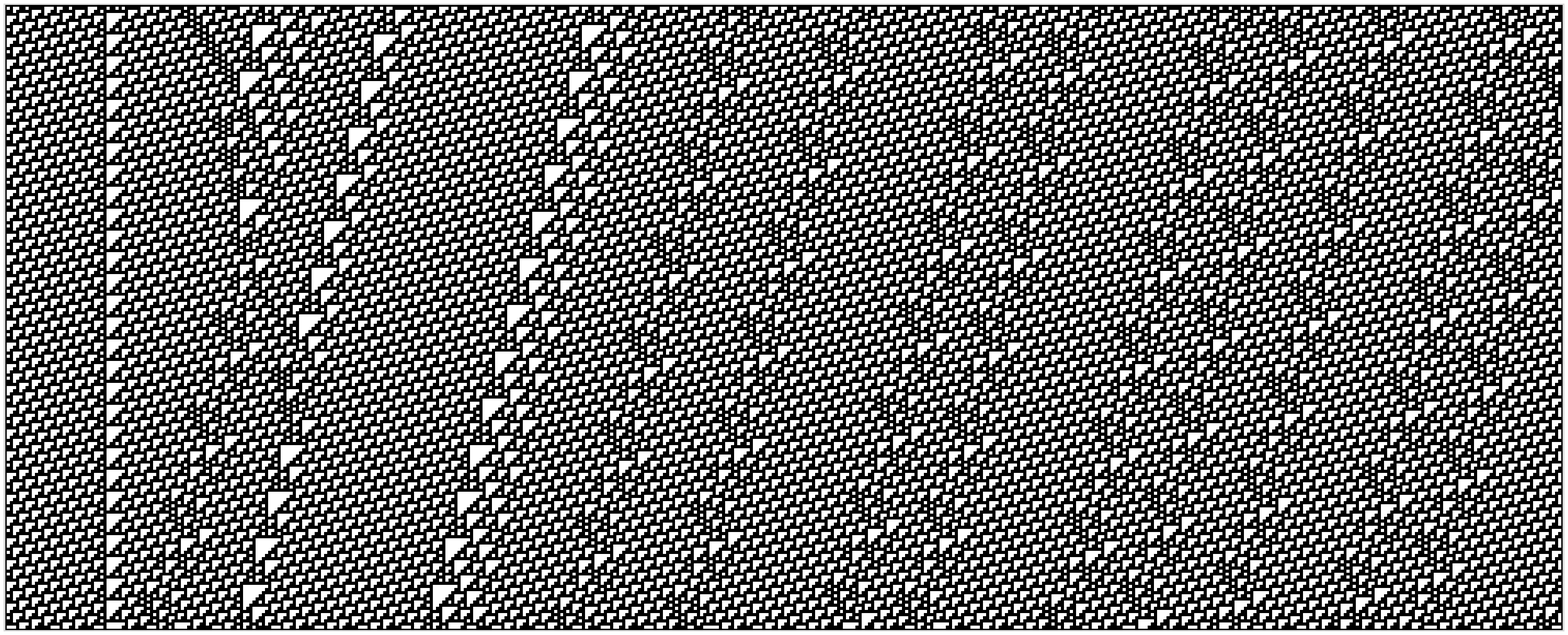}}
\caption{Space-time pattern of the leftmost part of section 16  in Fig.3 for the
 first 200 steps. Array size is 500.
Starting from the left, there are the rightmost part of tape data '1' (1Ele$\_C_2$), leader
 (SepInit$\_E\overline{E}$), and the left part of  table data '1' (1BloP$\_\overline{E}$).}
\label{space_time_sec16}
\end{figure}

The initial configuration in section 16 contains several patterns such as tape data, table data
or leader that separates packages of table data (Fig.~\ref{space_time_sec16}).
The group of these complicated structures makes the complexity high in this section. As time goes
by, however, these patterns move to the left and there remains the periodic background called
ether that causes the low value of complexity.

\begin{figure}
\scalebox{0.62}{\includegraphics{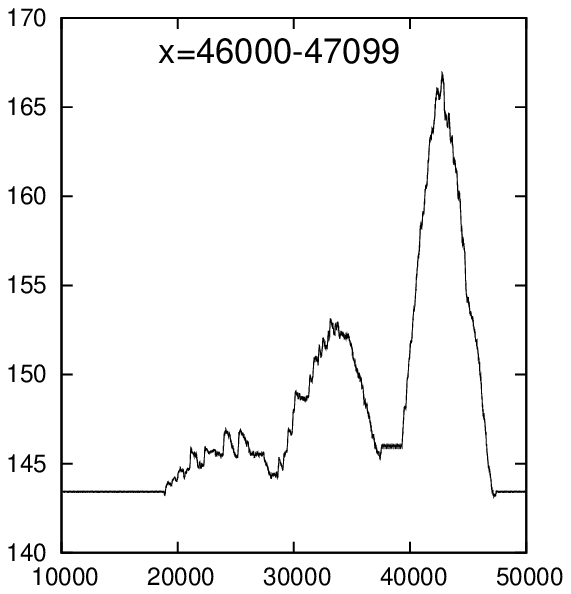}}
\scalebox{0.62}{\includegraphics{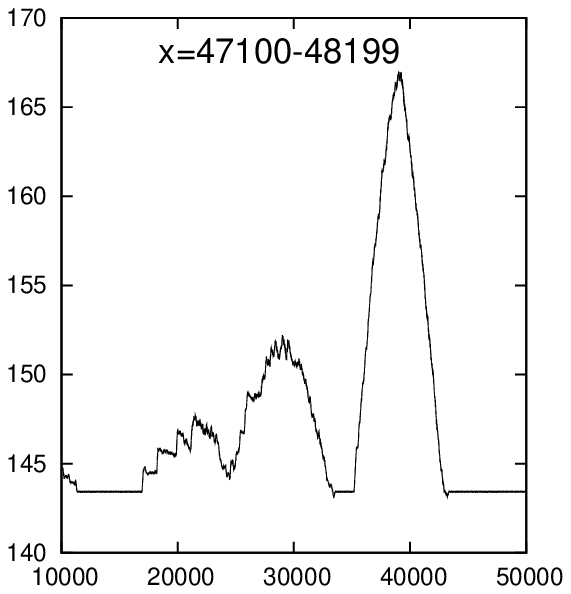}}
\scalebox{0.62}{\includegraphics{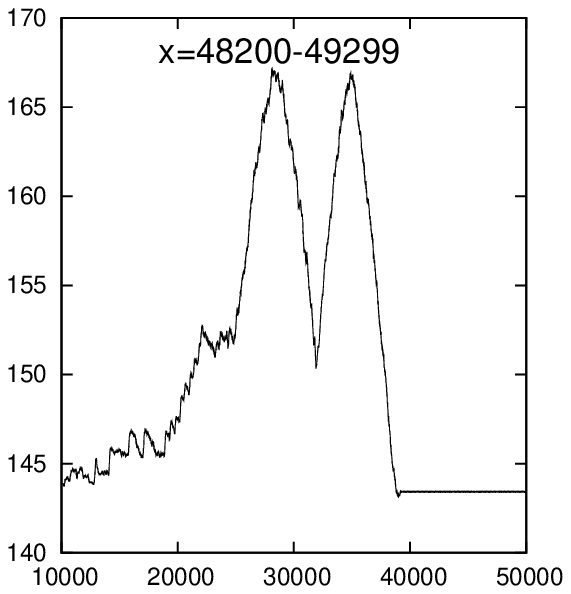}}
\caption{Moving average of the LZ complexity in the three parts of the section 14  in Fig.3 during time step from $t=10,000$ to $t=50,000$.}
\label{moving_avg_sec14}
\end{figure}

\begin{figure}
\scalebox{0.38}{\includegraphics{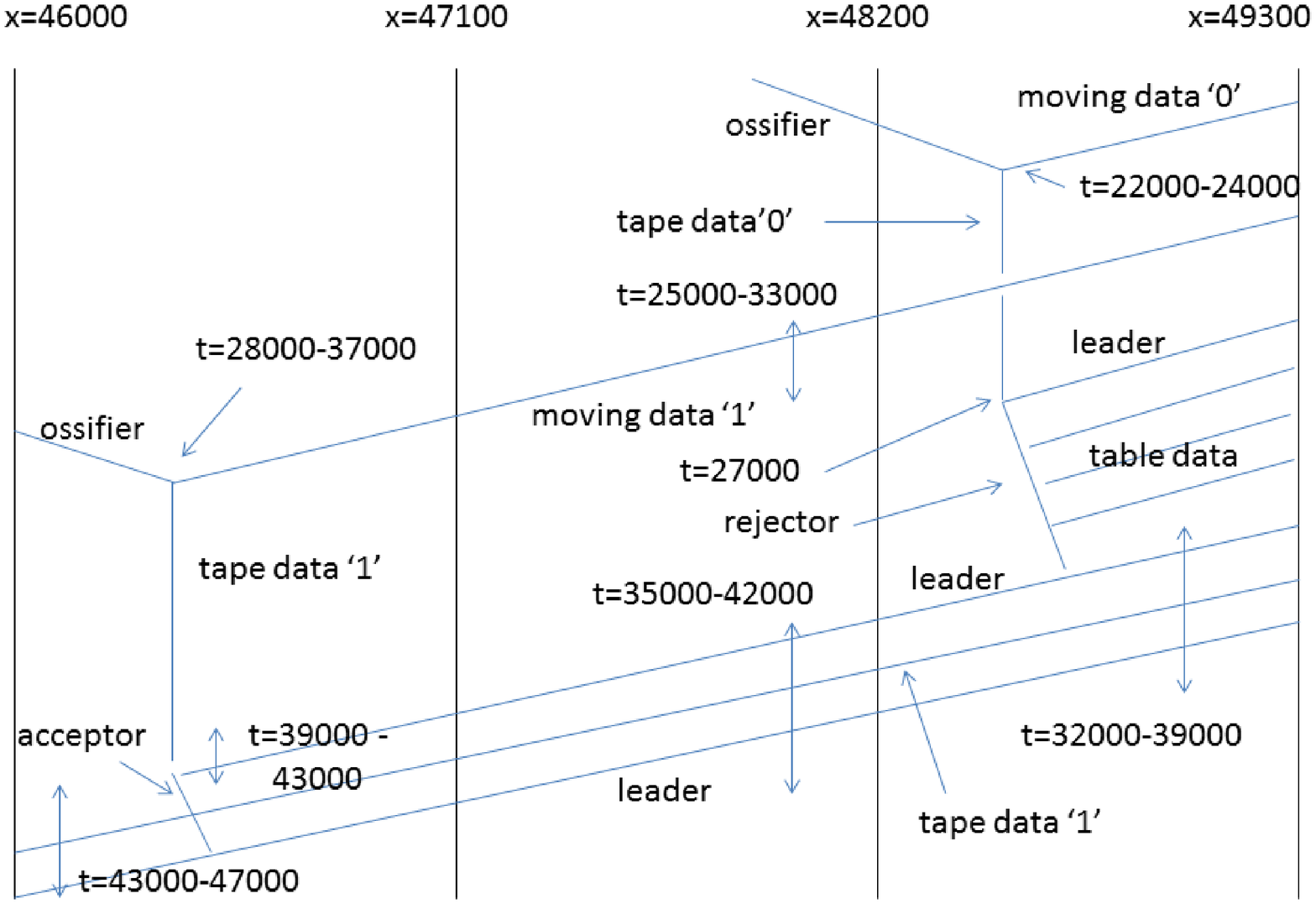}}
\caption{Diagram of patterns in section 14 in Fig.3. Time goes from top to bottom.}
\label{diagram_sec14}
\end{figure}

\begin{figure}
\scalebox{0.34}{\includegraphics{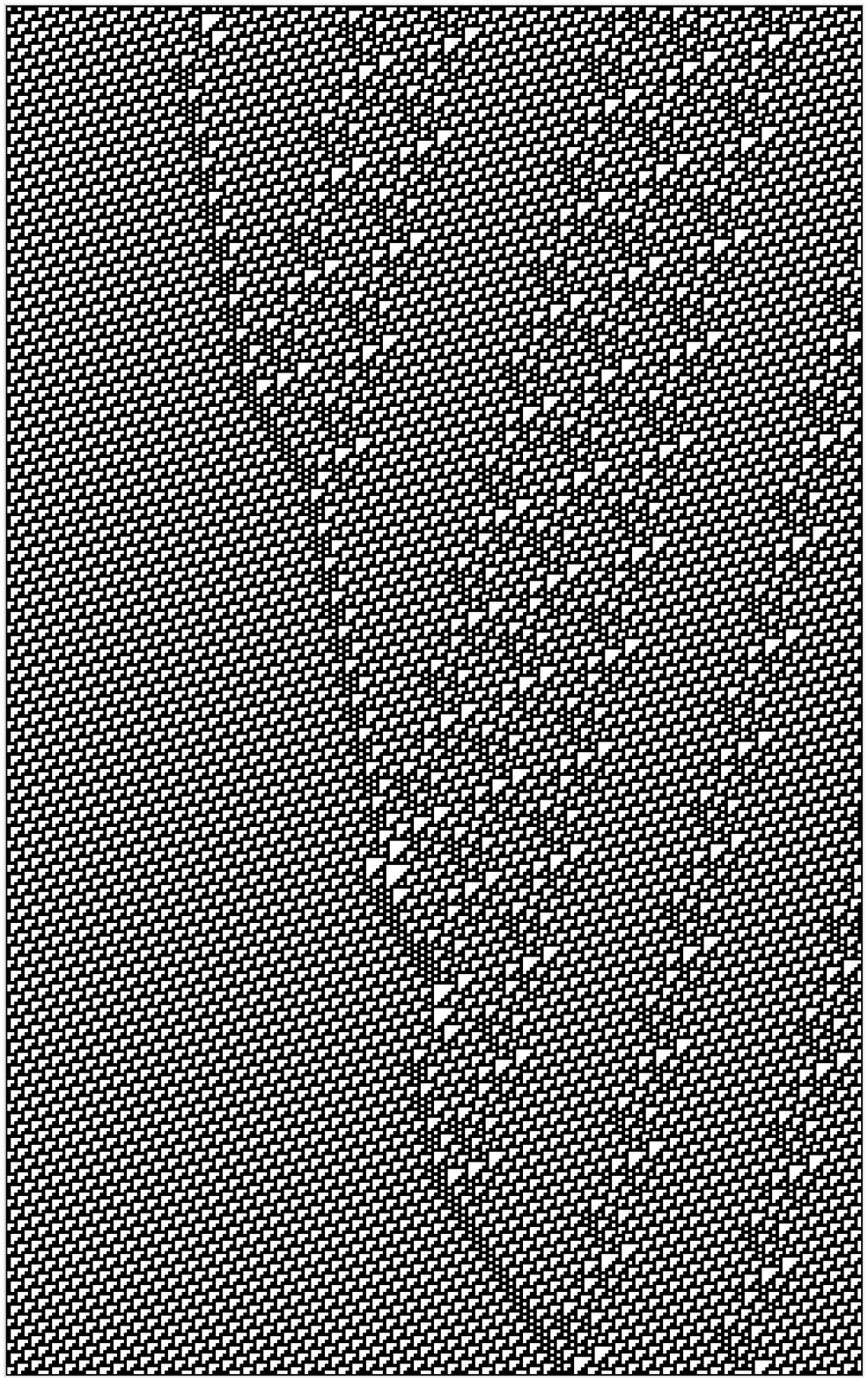}}
\scalebox{0.34}{\includegraphics{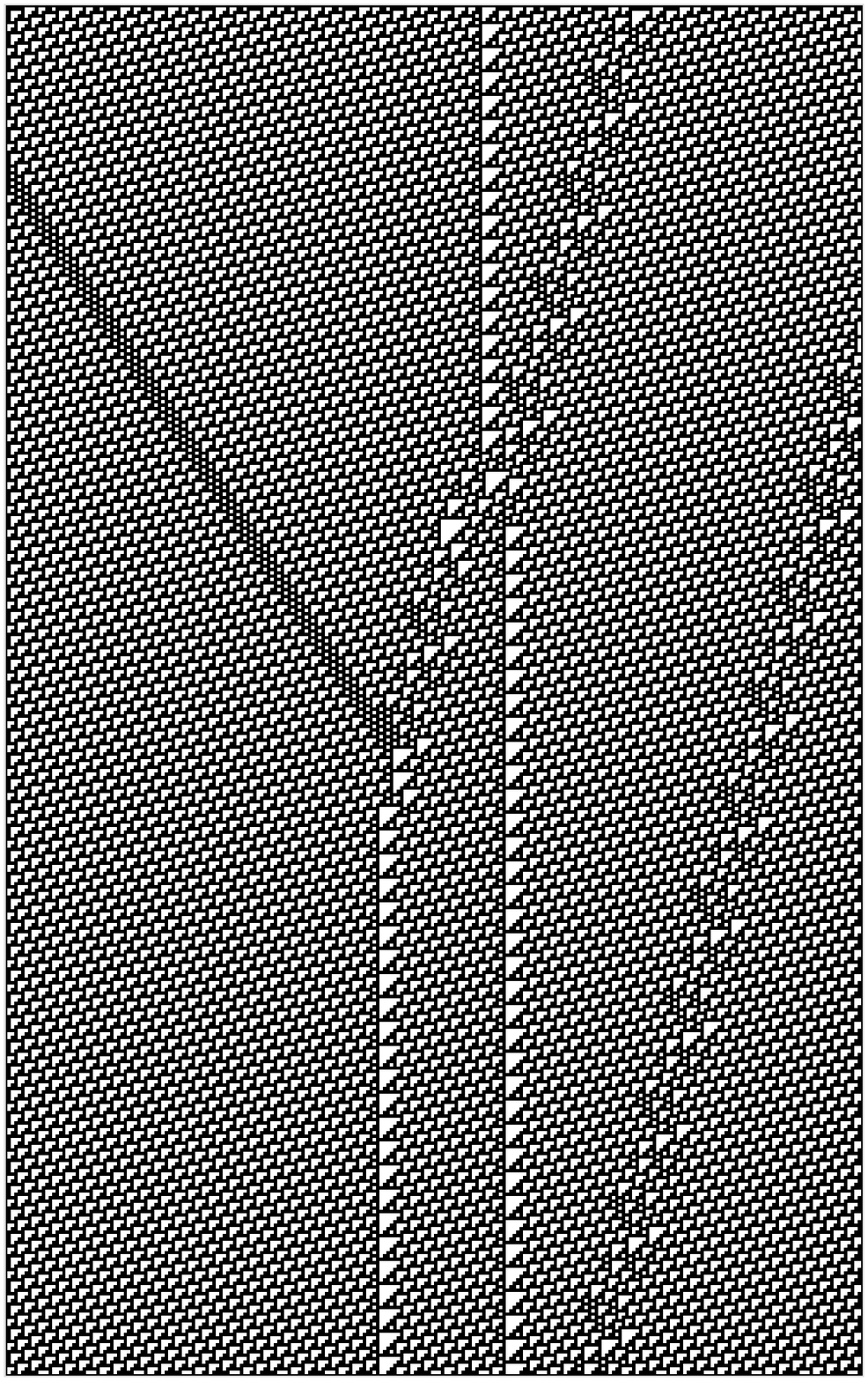}}
\caption{Space-time pattern of the part of section 14 in Fig.3.
Left: A rejector is erasing table data. Right: Collision between an
 ossifier moving from the left
 and moving data from the right is constructing a table data. Both
 pictures show 400 steps of evolution of the area of 250 cells.}
\label{space_time_sec14}
\end{figure}

For the detailed analysis of section 14, we divide it into three parts of array size 1100 and calculated the moving average
(period:100) of LZ complexity for each part during time step from $t=10,000$ to $t=50,000$ as shown in Fig.~\ref{moving_avg_sec14}.
The explanation in this paragraph is shown diagrammatically in Fig.~\ref{diagram_sec14}.
The right of Fig.~\ref{moving_avg_sec14} shows LZ complexity in the part of $x=48,200 \sim 49,299$ (the index of the leftmost cell of the array is given by $x=0$).
As a moving data '0' (0Add$\_\overline{E}$) is coming from the right, the LZ complexity starts increasing from $t=20,000$.
While the collision between the moving data '0' and an ossifier creates a tape data '0' (0Ele$\_C_2$) from $t=22,000$ to $t=24,000$ , the  LZ complexity 
does not vary a lot because tape data do not move.
The LZ complexity increases from $t=25,000$ as a leader and three table data come from the left.  When the leader collides with a tape data '0' at about $t=27,000$,
the LZ complexity reaches a maximum. While a rejector created by the collision is erasing table data as show in the left of Fig.~\ref{space_time_sec14}, the LZ complexity decreases a lot from $t=28,000$ to $t=32,000$.

When the moving data '1' created in section 15 pass through the part of  $x=47,100 \sim 48,199$ to the left, the LZ complexity temporarily increase from $t=25,000$ to $33,000$ as shown in the middle of Fig.~\ref{moving_avg_sec14}. Finally the moving data '1' collides with an ossifier coming from the left and converts into tape data '1' as shown in the right of Fig.~\ref{moving_avg_sec14}. That corresponds to the bump at $t=28,000 \sim 37,000$ in the part of $x=46,000 \sim 47,099$ shown in the left of Fig.~\ref{moving_avg_sec14}.
The sharp increase from $t=32,000$ in the right of Fig.~\ref{moving_avg_sec14} is caused by leaders and table data moving from the right. These structures bring a same result from $t=35,000$ in the middle of Fig.~\ref{moving_avg_sec14} and from $t=39,000$ in the left of Fig.~\ref{moving_avg_sec14}, as they move to the left. 
They collide with tape data '1' and convert into a moving data '1' that goes away to the left. That causes the sharp decrease from $t=43,000$ to $t=47,000$ in the left of Fig.~\ref{moving_avg_sec14}. Since there remains only ether after $t=47,000$, the value of LZ complexity is low.

\section{Discussion}
\begin{figure}
\scalebox{0.34}{\includegraphics{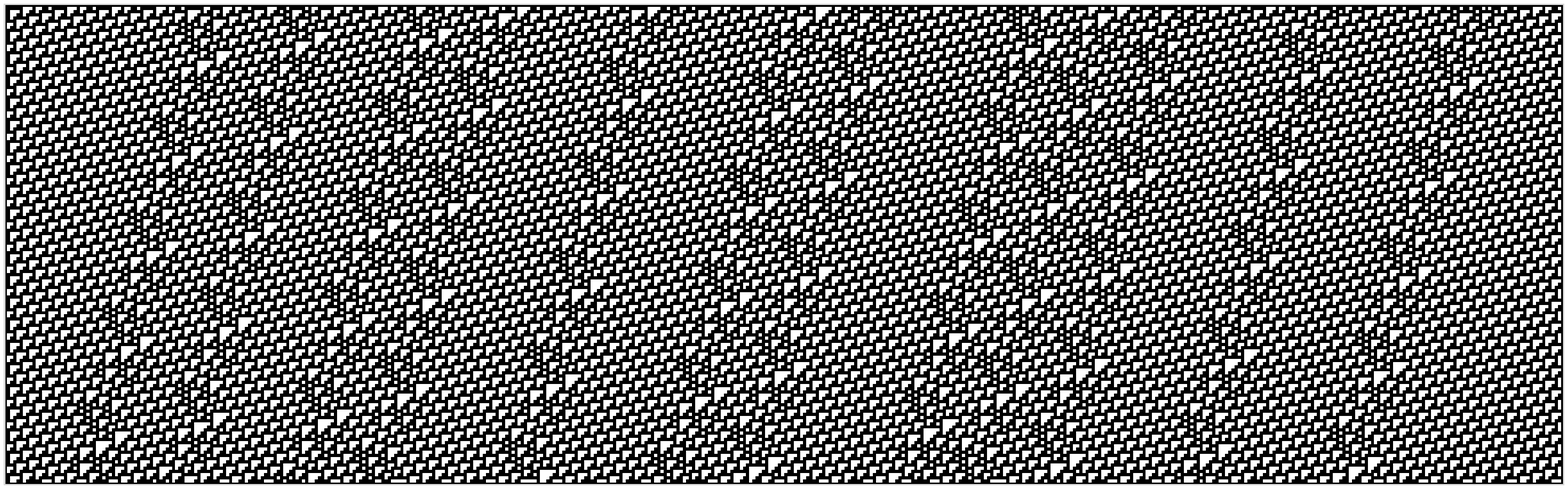}}
\scalebox{0.34}{\includegraphics{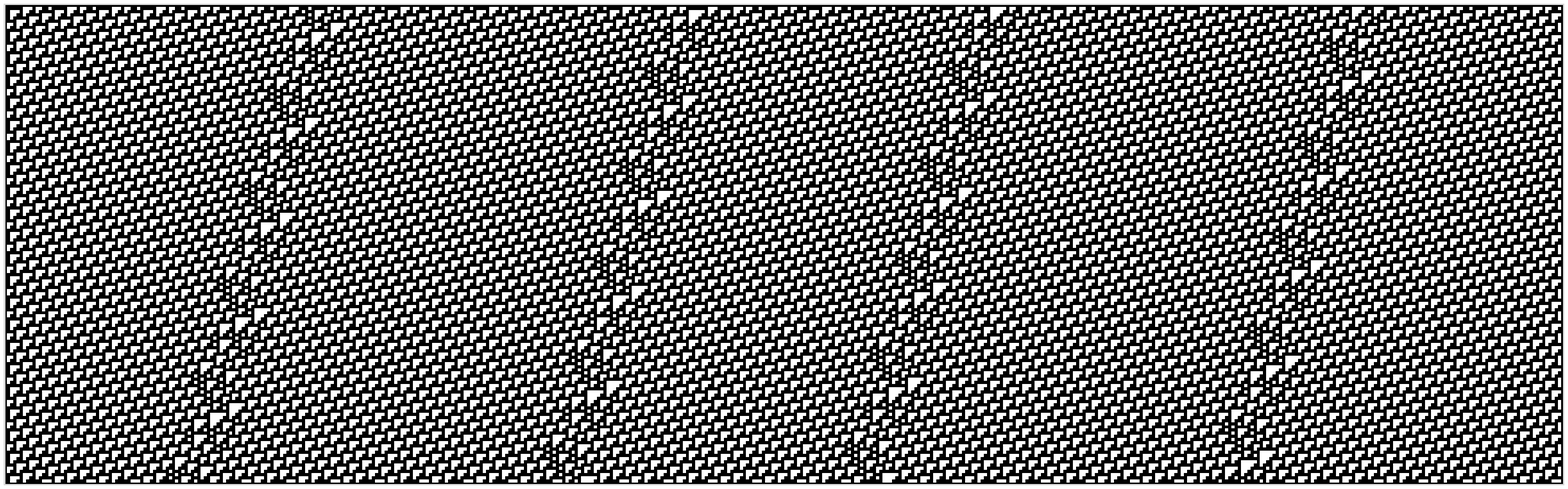}}
\caption{Space-time pattern of table data '1' (top) and moving data
 '1' (bottom). The array size is 490 in both cases.}
\label{tabledata_movingdata}
\end{figure}

As we explained in the previous section, by dividing the whole array into small parts, we can correspond the change of LZ complexity to the events occurring on the array such as incoming of propagating patterns, collision between several patterns, and outgoing of them.
In particular the significant decline of LZ complexity is caused by a transformation from table data to moving data or an elimination of table data. Figure~\ref{tabledata_movingdata} shows the space-time pattern of table data '1' (1BloP$\_\overline{E}$) (top) and moving data '1' (1Add$\_\overline{E}$) (bottom). It is apparent that the pattern of table data '1' is more complicated than that of moving data '1'.
Therefore the transformation from table data to moving data brings about the decline of LZ complexity. And the elimination of table data by a rejector has the same result. It seems that the decline of LZ complexity observed in Fig.~\ref{moving_avg} is caused by the events of either of these two types.

In this research we employed the periodic boundary conditions. The ossifiers, the table data, and the leaders are built in advance in the initial configuration. They are consumed during the evolution and are not supplied from the outside.
It is uncertain about how LZ complexity varies in time if they are supplied regularly and eternally.
We are planning to employ "discharging" boundary conditions that can supply these patterns from the outside.

%
%

\end{document}